# Trans GAN-WT: A Feature Extraction and Interactive Learning-Based Anomaly Detection Model for Wind Turbine Time Series Data

Jingzhe Kang[1,*]

[1]School of Mathematics, Shandong University, Jinan 250100, China
Email: kangjingzhe@mail.sdu.edu.cn

**Abstract:** With the increasing scale and number of wind farms, wind turbines' daily operation and maintenance costs are increasing. To reduce operation and maintenance costs and enhance the reliability of wind turbine and system operation data before reaching catastrophic failures, monitoring the operating status of the equipment and detecting failures at an early stage is crucial. It is of great practical significance to utilize the working condition data for abnormal assessment of the operating status of wind turbines to realize abnormal monitoring of the operating status of wind turbines. However, the existing anomaly detection methods can neither perform effective relational modeling in data filled with a large amount of redundant information nor reasonably utilize the valuable anomaly data. For this reason, this paper proposes an anomaly detection model that fuses a Transformer and a generative adversarial network. Firstly, it reduces the leakage detection rate of minor deviation anomalies by amplifying the reconstruction error. Secondly, it uses autoregressive inference to extract multimodal features to enhance the stability and generalization ability of training. Finally, the temporal feature extraction module is constructed to promote the interactive learning between features of different time scales and effectively reduce the time redundancy. The results of multiple sets of experiments conducted on real WTG datasets show that TransGAN-WT achieves an average F1 score of 96.10% across multiple wind turbine datasets, which is 5.84% and 2.89% higher than several other state-of-the-art baseline methods. It also realizes a false positive rate (FPR) of 0.06%, and is verified by the Wilcoxon signed-rank test to have achieved a statistically significant performance enhancement compared to the state-of-the-art baseline methods, effectively ensuring the stable operation of wind turbines.

## 1. INTRODUCTION

The problem of energy shortage is the primary problem that limits the international community from maintaining rapid development today [1]. As a renewable and sustainable clean energy source with wide geographical distribution and colossal development potential, wind energy is an essential part of the energy system of today's international society. Currently, wind power generation is in a period of rapid development, and the scale of wind farms continues to expand. However, modern wind turbines are highly costly to manufacture and install on-site [2] as a large and highly complex electrical system. Due to the distribution characteristics of wind energy resources in the geographical environment and the constraints of power grids, life, and transportation, most of the wind farm sites are located in remote mountainous areas, near the sea, and the offshore regions with harsh natural environments (e.g., high temperatures, humidity, salinity, and strong winds) [3]. Continuously working under such variable and harsh environments for an extended period can easily lead to early degradation and failure of the working parts of wind turbines [4]. Considering the geographic characteristics of the wind farm site, it will cost a considerable amount of time and money when the wind turbines have significant failures and require substantial repairs [5]. Therefore, detecting and discovering data anomalies for the early stage of failure is very important to improve the reliability of wind turbines [6]. Continuous monitoring of the equipment health and operational status of wind turbines using anomaly detection methods is essential to effectively reduce the cost of routine operation and maintenance of wind turbines and increase the reliability of wind turbines before they reach catastrophic failure.

Supervisory control and data acquisition (SCADA) systems installed in wind farms record a large amount of system data and equipment operation data collected by sensors of wind turbine components, and the SCADA data of wind turbines are typical multivariate time series (MTS). Series (MTS), and these time series variables reflect the state trajectories and potential correlations of different wind turbine components [7]. Anomaly assessment of wind turbine operation status on SCADA system data can be realized to monitor wind turbines' health status to detect abnormal equipment operation in time [8]. Abnormality detection can identify and discover the root cause of abnormal wind turbine operation by searching for abnormal data in the SCADA system data through abnormal assessment of wind turbine operation status.

However, current anomaly detection techniques are still unable to meet the demand for long-term stable operation and more accurate detection of wind farm systems. Specific challenges in the existing detection tasks include: (1) The category share of the power unit operating condition data samples is seriously unbalanced. Meanwhile, the anomaly information may become more complicated in sensor measurement, data conversion, transmission, and storage. How to effectively discover and expand the anomaly features in the limited anomaly data is a crucial issue in anomaly detection. (2) The data collection process for wind turbines involves significant redundancy. Due to minimal changes in environmental conditions (such as wind speed and temperature) during specific periods, the collected data may remain stable over time, resulting in substantial temporal redundancy. In such a complex and variable turbine environment, detecting anomalies involves handling a large volume of data collection and analysis. Both insufficient and excessive data can impact the effective detection of anomalies.

Researchers have constructed many wind turbine operation state abnormality assessment modeling methods based on SCADA system data, which can be roughly divided into two categories: one based on equipment modeling methods and the other based on data-driven methods. The method based on the equipment model is mainly through the distribution characteristics of the data combined with the a priori theoretical knowledge, including the physical processes in the electrical system of the wind turbine, the interaction between the critical sub-equipment, and using mathematical statistics to carry out the abnormal assessment of the equipment operating state. Zhang et al. constructed a relationship model between the wind speed and the power through the local outlier algorithm. Then, the abnormal operating state is assessed by analyzing the relative data density of data near the model. Density, the abnormal operation state, is estimated [9].

Zhang Fan et al. [10] established an input-output relational model based on SCADA system data using a polynomial regression fitting technique from the unit's operating characteristics. They used this model to identify abnormal information in the data from actual equipment. Wang Xin et al. [11] classified the wind speed power data based on the distribution characteristics of the wind power curve. They used a sub-area algorithm to evaluate the anomalous data. However, these methods did not fully consider the temporal features of SCADA system data during the modeling process, leading to difficulties in accurately capturing anomalies caused by wind turbine operating status changes. In addition, the complex physical and mathematical models used in these methods may be too demanding on the system's overall performance, and it is also challenging to model nonlinear sequences.

It is worth noting that anomalies in wind turbines are not only manifested as deviations in operational parameters but may also stem from structural damage and degradation. From the perspective of structural health monitoring, real-time damage identification technology offers another crucial approach for detecting anomalies in wind turbines. The high-order eigenvalue perturbation framework enables real-time damage identification through stiffness matrix decomposition and natural frequency tracking, capturing system characteristic changes caused by physical damage, such as fatigue and cracks, at the structural dynamics level. This approach demonstrates unique advantages in wind turbine shutdown detection and structural integrity assessment [12]. This method is based on modal analysis theory. It identifies early damage by monitoring subtle changes in modal parameters, such as structural natural frequencies and mode shapes, which possess clear physical meaning and good interpretability. However, such methods primarily focus on identifying structural anomalies, with relatively limited detection capabilities for non-structural issues such as electrical system faults and control strategy abnormalities. Meanwhile, in complex wind farm environments, environmental factors such as temperature changes and foundation settlement can also affect the dynamic characteristics of structures, potentially interfering with damage identification based on natural frequencies.

In contrast, data-driven methods provide new insights for addressing the limitations of physics-based model approaches by directly learning anomaly patterns from massive operational data in SCADA systems. These methods extract the variation patterns of various performance parameters from historical data. Compared with the detection method based on the equipment model, it requires less information about the equipment domain, has relatively low hardware, and, at the same time, can deal with the nonlinear relationship well. Liu et al. utilized Support Vector Regression (SVR) [13] to map the nonlinear relationship between the anomaly threshold and the device operation state and realized the detection method of the adaptive threshold. Still, the method relies too much on selecting the kernel function in the middle. Chen et al. [14] established a Stacked Denoised Autoencoder (SDAE) model based on an optimized sliding window. They used the kernel density estimation method to fit the probability density distribution of the data monitoring indexes of wind turbines under normal operating conditions. Still, the length of the sliding window interval of this method primarily affects the detection accuracy.

Yu et al. [15] established an energy efficiency state indicator system based on energy flow, using temperature parameters as the basis, and employed an autoregressive integrated moving average model-support vector machine (ARIMA-SVM) combined model to achieve time series prediction of energy efficiency. However, this approach requires determining the energy efficiency anomaly index values based on historical experience. Wang et al. [16] employed a residual sequence detection strategy based on Mahalanobis distance to achieve high-precision anomaly identification for wind turbine blades. Chen et al.[17] proposed an intelligent anomaly state determination method for wind turbine SCADA data based on a time-varying hidden Markov model, which can detect anomalies by comparing the properties of normal data with those of classified sub-datasets. However, this method is sensitive to the selection of initial states, and inappropriate initial states may lead to slow model convergence or convergence to local optima.

Li et al. [18] proposed a wind turbine SCADA anomaly data identification method based on the density-based spatial clustering of applications with noise (DBSCAN) model. By introducing prediction error and classification accuracy and selecting optimal clustering parameters, including the neighborhood radius and the minimum number of neighboring sample points, the method combines different data features to identify anomaly data. Nevertheless, suppose the original data contains substantial noise or exhibits an imbalance of positive and negative samples. In that case, these characteristics may be incorrectly interpreted as normal distributional properties of the data, thereby affecting the accuracy of clustering results and anomaly detection.

Kini et al.[19] proposed a semi-supervised data-driven monitoring technique based on Dynamic Independent Component Analysis (DICA), which employs kernel density estimation to establish nonparametric detection thresholds, but the existing methods still have challenges in dealing with the temporal dynamic characteristics and non-Gaussian features of wind turbine SCADA data. Harrou et al. [20] proposed a semi-supervised data-driven method based on Independent Component Analysis (ICA) and Kantorovich Distance (KD), which can efficiently handle multivariate non-Gaussian data and compute detection thresholds through kernel density estimation. However, existing sensor fault detection methods still face challenges in handling multivariate non-Gaussian data, sensitivity to slight variations, and setting appropriate detection thresholds to avoid false alarms, especially for large-scale wind turbines, where constructing analytical monitoring models becomes particularly complex and time-consuming.

Although the aforementioned data-driven methods have demonstrated particular effectiveness in wind turbine anomaly detection, most of them still rely on point predictions or fixed-threshold strategies, which struggle to capture the dynamic uncertainty boundaries of normal operating conditions adequately. Recently, probabilistic density forecasting has emerged as a key approach to enhancing both the robustness and interpretability of anomaly detection systems. For example, Lu et al. [21] proposed a non-crossing sparse-group Lasso quantile regression deep neural network (SGLQRDNN) for electricity load forecasting. By jointly estimating multiple quantiles under non-crossing constraints, their model ensures the monotonicity and validity of the resulting prediction intervals. Moreover, the incorporation of a sparse-group Lasso regularization strategy simultaneously enables network sparsification and selection of critical features, thereby reducing structural complexity. This method not only achieves high-accuracy probabilistic forecasting but also significantly improves training efficiency and generalization performance. Although applied in the context of load forecasting, the underlying principles—uncertainty quantification, adaptive confidence interval construction, and model simplification—provide valuable insights for wind turbine anomaly detection, particularly in the design of dynamic thresholds and modeling of residual distributions.

In summary, data-driven methods exhibit high dependence on model parameters and historical experience, with limited adaptive capability for detection tasks under different scenarios. When confronted with poor data quality, particularly data with insignificant variation ranges and imbalanced positive and negative samples, traditional detection algorithms still present numerous shortcomings.

With the rapid development of neural network technology, it has been widely adopted as an end-to-end computational method in power systems, equipment detection, and fault alarms, among other applications. Xu et al. [22] proposed a novel adaptive fault detection scheme that integrates Random Forest (RF) with an adaptive Cumulative Sum (CUSUM) technique. The adaptive CUSUM represents an improvement over conventional alarm rules for wind turbines (WTs): rather than triggering an alert immediately when a single outlier exceeds a fixed threshold, it monitors the persistent degradation of WT performance and automatically resets after an anomaly is detected. This approach significantly enhances the accuracy of fault feature extraction, thereby reducing false alarms and enabling fully automated fault detection. Guo et al. [23] map data to the output layer and maintain the original topology using Self-Organizing Maps (SOM) for the device state anomaly monitoring mechanism. Wang et al.[24] designed multiple CNN modules with different-sized convolutional kernels to perform abnormal state monitoring of wind turbines to extract correlations between multiple sensor variables and temporal dependencies hidden in different variables in parallel.

Long Short-Term Memory (LSTM) neural networks, as an improved variant of RNN, incorporate long-term dependencies between time series data and nonlinear characteristics among data into model training through their distinctive gating structure. Zhu et al. [25] applied convolutional neural networks (CNN) and long short-term memory networks (LSTM) to the condition monitoring of wind turbine gearbox bearings for the identification of impending failures. On this basis, Liu et al. [26] proposed an anomaly assessment method of wind turbine operation state based on LSTM-AE, which can effectively process multidimensional time series data, extract the hidden nonlinear relationships, and obtain the equipment performance indexes through the calculation of the error between output reconstructed values and the original values. It is worth noting that it is difficult for LSTM-based models to capture the redundancy information at all time steps, mainly when the redundancy in the sequence is distributed over different time scales and if the redundancy information is very complex or occurs in a nonlinear manner, the LSTM may need a deeper network or more parameters to learn these patterns.

Despite the significant progress made by those mentioned above, neural network-based methods in wind turbine anomaly detection still face numerous challenges in practical applications. Particularly when dealing with small-deviation anomalies and temporal redundancy issues commonly present in wind turbine SCADA data, the limitations of existing methods become increasingly apparent. Therefore, it is necessary to conduct a systematic review and comparative analysis of existing wind turbine anomaly detection methods to clarify the advantages and limitations of various approaches. Table 1 provides a systematic summary of existing wind turbine anomaly detection methods and conducts an in-depth comparative analysis from three key dimensions: interpretability, equipment dependency, and parameter sensitivity, as detailed in Table 1.

**Table 1 Comparison Of Wind Turbine Anomaly Detection Methods**

| Method category | Method | Fundamental Technology | Advantages | limitations | Explainability | | Equipment Dependency | | Parameter Dependency | |
|---|---|---|---|---|---|---|---|---|---|---|
| | | | | | low | high | low | high | low | high |
| Device-based approach | Local Outlier Factor Algorithm | Construct a wind speed-power relationship model and analyze the relative density of data. | • Identify power curve anomalies based on physical relationships and prior theoretical knowledge | Difficult to capture trends in operational status changes | | ✓ | | ✓ | | ✓ |
| | Polynomial regression fitting | Establish an input-output relationship model based on operating conditions | • Applicable to multivariate analysis, requiring the establishment of a precise input-output mapping | Complex physical mathematical models need to be established | | ✓ | ✓ | | | ✓ |
| | Regional algorithm | Classify data based on the characteristics of the wind power curve distribution | • Classify and process different operating conditions based on wind power curve characteristics | • High modeling complexity<br>• Insufficient consideration of dynamic characteristics | | ✓ | | ✓ | | ✓ |
| Based on structural health monitoring | High-order eigenvalue perturbation framework | Stiffness matrix decomposition and natural frequency | Detection from the structural dynamics perspective• Suitable for early | •Mainly targets structural abnormalities; limited ability to | | ✓ | ✓ | | | ✓ |
| methods | | tracking | damage detection, with real-time damage localization capabilities | detect non-structural problems. | | | | | | |
| Based on data-driven methods | Support Vector Regression (SVR) | Nonlinear relationship between mapping anomaly thresholds and operating status | • Strong ability to handle nonlinear relationships, adaptive threshold setting | • Over-reliance on kernel function selection, sensitive to sample quality | | ✓ | ✓ | | | ✓ |
| | Stack Noise Reduction Autoencoder SDAE | Based on an optimized sliding window and kernel density estimation | • Can process noisy data and perform automatic feature extraction. | • Sliding window length affects detection accuracy | ✓ | | ✓ | | ✓ | |
| | ARIMA-SVM combination model | Energy efficiency status index system based on energy flow | • Combining time series to consider changes in energy efficiency status, with the advantage of multi-model fusion | • Historical experience is needed to determine the anomaly index. | | ✓ | ✓ | | ✓ | |
| | Mahalanobis distance residual detection | Self-supervised method based on multi-channel vibration | • High-precision anomaly detection, combined with multi-channel | • Mainly targets leaf abnormalities, with relatively | | ✓ | | ✓ | | ✓ |

| | | | | | | | | |
|---|---|---|---|---|---|---|---|---|
| | principal component features | information fusion | limited application scenarios | | | | | |
| Time-varying hidden Markov model | Compare normal data attributes and sub-dataset attributes | • Handling time-varying characteristics • Intelligent status determination • Probability models are highly reliable | • Sensitive to initial state selection | ✓ | | ✓ | | ✓ |
| A non-crossing sparse-group Lasso-quantile regression deep neural network (SGLQRDNN) | By jointly estimating multiple quantiles under non-crossing constraints to achieve probabilistic density forecasting, and integrating sparse-group Lasso regularization to accomplish network compression and critical feature selection simultaneously. | • It effectively enhances model interpretability and reduces the risk of overfitting.<br>• Compared to the traditional approach of independently training each quantile separately, it achieves a better balance between efficiency and performance. | • The method has not yet been comprehensively extended or validated for adaptability to the dynamic environmental conditions commonly associated with wind turbine anomalies. | ✓ | ✓ | | ✓ | |

| | | | | | | | | | |
|---|---|---|---|---|---|---|---|---|---|
| An adaptive fault detection scheme combining Random Forest (RF) and adaptive Cumulative Sum (CUSUM). | Combining Random Forest with a time-varying adaptive CUSUM chart enables sensitive, robust, and automated detection of subtle residual changes. | Significantly reduces false triggers caused by transient noise or short-term disturbances. | Its response to abrupt faults may be delayed. | | ✓ | ✓ | | ✓ | |
| DBSCAN clustering | Density-based spatial clustering combined with prediction error | Multi-feature fusion judgment, no need to preset the number of clusters | • High data quality requirements, requiring a balanced distribution of positive and negative samples | | ✓ | ✓ | | | ✓ |
| Dynamic Independent Component Analysis (DICA) | Combining $I^2d$, $I^2e$, and SPE metrics with DEWMA charts, kernel density estimation thresholds | High sensitivity to detecting minor faults, with minimal labeling requirements. | Insufficient ability to model complex nonlinear relationships | | ✓ | ✓ | | | ✓ |
| ICA-Kantorovich | Independent component | • Effectively handles | • Limited sensitivity to | ✓ | | ✓ | | ✓ | |

| | | | | | | | | | | |
|---|---|---|---|---|---|---|---|---|---|---|
| | distance (KD) | analysis combined with KD distance measurement, kernel density estimation threshold | multivariate non-Gaussian data with flexible threshold settings | small changes | | | | | | |
| | Self-Organizing Map (SOM) | Data is mapped to the output layer and retains its original topological structure | • Maintains original data topology relationships and powerful visualization capabilities | Lack of deep feature extraction | ✓ | | | ✓ | | ✓ |
| | Multi-scale CNN | Parallel feature extraction using multiple CNN modules with different-sized convolution kernels | • Multi-perspective spatiotemporal feature fusion and parallel extraction of sensor variable correlations | Requires a large amount of training data | ✓ | | | ✓ | ✓ | |
| Based on end-to-end detection methods | CNN-LSTM hybrid model | Convolutional neural networks combined with long short-term memory networks | • Deep feature extraction using CNN spatial features and LSTM temporal features | • The model structure is complex, training time is long, and data quality requirements are high | ✓ | | | ✓ | | ✓ |
| | LSTM | LSTM-based | • Effectively handle | • Difficult to | | ✓ | | ✓ | ✓ | |

| | | | |
|---|---|---|---|
| Autoencoder (LSTM-AE) | autoencoder reconstruction error detection | multidimensional time series data and uncover hidden nonlinear relationships | capture redundant information at different time scales; complex redundant patterns require deeper networks |

The motivation of this paper is to solve the problems of slight data deviation and time redundancy in SCADA data for wind turbine anomaly detection and to ensure the safety of wind turbines during the project's life cycle. We propose an anomaly detection model that combines Transformers and Generative Adversarial Networks (GANs) to address this. By amplifying reconstruction errors, the model mitigates the issue of missing minor deviation anomalies and employs autoregressive inference to extract multimodal features, enhancing the stability and generalization of the training process. Additionally, we have designed a temporal feature extraction module to facilitate the interaction learning of features at different time scales and reduce temporal redundancy in time series data, providing a new approach for analyzing and detecting anomalies in wind turbine data.

In addition, this paper performs a comprehensive anomaly data detection task on the selected wind turbine data by using the wind turbine operation data of a real-world wind farm as the research object. The main contributions of this paper can be summarized as follows:

We propose Trans GAN-WT, an anomaly detection model that effectively fuses Transformers and Generative Adversarial Networks (GANs). In contrast to existing methods that merely stack or combine these components in a loosely coupled or parallel fashion—without establishing dynamic interaction between them—our approach introduces an elegant fusion mechanism [27][28]. Specifically, it fully leverages the self-attention capability of the Transformer for modeling complex temporal dependencies [29], while simultaneously exploiting the distributional characteristics of the GAN discriminator's output to implicitly learn an adaptive decision boundary through adversarial training. This significantly enhances sensitivity to subtle, persistent deviations. As a result, Trans GAN-WT fundamentally addresses the challenge of modeling temporal correlations in multivariate wind turbine time-series data and effectively overcomes the limitation of conventional Transformer-based encoder-decoder architectures, which often fail to detect anomalies when the deviation between normal and anomalous data is slight.

Trans GAN-WT utilizes self-attention mechanisms to separate complex component features within time series data and capture dependencies between different features. At the same time, it leverages Generative Adversarial Networks to amplify anomalous features, significantly enhancing anomaly detection accuracy.

To process temporal data more efficiently, Trans GAN-WT has designed a novel temporal dependency capturer, a module that can efficiently capture multi-scale temporal dependencies while reducing temporal redundancy. The catcher can efficiently learn the temporal information in the subsequence after down sampling and capture the dependencies with less redundancy.

A series of experiments on actual wind turbine datasets show that Trans GAN-WT outperforms the current state-of-the-art baseline methods regarding F1 values and realizes a significant improvement in detecting anomalous data in the wind turbine context.

The rest of the paper is organized as follows: Section 2 defines the problem associated with anomaly detection in wind turbines. Section 3 presents the details and implementation of Trans GAN-WT proposed in this paper. In Section 4, we conduct a series of experiments on the dataset to demonstrate the effectiveness of our proposed method by comparing it with existing models. Section 5 provides a comprehensive paper summary and an outlook for future work.

## 2. PROBLEM DEFINITION

### 2.1 Problem formulation

Wind turbine multivariate time series usually contain sequence data for multiple variables at the same sampling interval. Given a set of SCADA data $X \in R^{M \times T}$, where $M$ and $T$ denote the number of variables and the time series length, respectively. The input data $X$ is first preprocessed and processed by the sliding window technique after the anomaly detection model completes the training and detection model to obtain the state of the data points $\{y_1, y_2, ..., y_t\}$ , which $y_t \in \{0,1\}$ serves as the moment $t$ whether the data is abnormal or not, $y=1$ is indicated as an abnormal point, and $y=0$ is a typical point.

Definition 1 (Anomaly Score): During the testing phase, the inference process is divided into two stages. The input window

$\hat{W}$ is sequentially reconstructed as $O_1$ and $O_2$ Based on the differences between the reconstructed vectors and the input window, the anomaly score is defined as given in Eq. (1).

$$s = \frac{1}{2} \| \mathbf{O}_1 - \hat{\mathbf{W}} \|_2 + \frac{1}{2} \| \hat{\mathbf{O}}_2 - \hat{\mathbf{W}} \|_2 \quad (1)$$

Specifically, given a segment of input data, the model is first trained on the training set. During the inference phase, unseen test data with the same sequence length as the training samples are divided into sliding windows and reconstructed sequentially. The anomaly score is then obtained by calculating the deviation between the input window $\hat{W}$ and its reconstructed vector. To achieve adaptive detection, the Peaks Over Threshold (POT) method is employed to generate the threshold dynamically. When the anomaly score at a given timestamp exceeds this threshold, the corresponding data point is identified as anomalous, resulting in the final predicted label set for the test sequence.

### 2.2 Data preprocessing

In wind turbine SCADA systems, the collected data often contains noise and outliers due to limitations in sensor accuracy, environmental factors, or communication issues. Sliding window techniques can effectively remove noise, smooth data curves, and enhance data reliability and accuracy, facilitating the more accurate identification of genuine anomalous data in subsequent processing. Meanwhile, sliding window techniques possess high computational efficiency and can effectively handle large-scale wind turbine datasets. When applying sliding window smoothing techniques to the original time series data $X = \{x_1^{'}, x_2^{'}, ..., x_T^{'}\}$, for each time point $t^{'}$, a sliding window of length $R$ is defined, and the average value of data points within this window is calculated as the smoothed data point $x_t$ :

$$x_t = \frac{1}{R} \sum_{i-\max(1,t^{'}-R+1)}^{t^{'}} x_i^{'} \tag{2}$$

For the time point $t^{'} < R$ , the window starts from $x_1^{'}$ and calculates the average value of the preceding $t^{'}$ points. For time point $t^{'} \geq R$ , the window contains $\{x_{t^{'}-R+1}, ..., x_t^{'}\}$ .

To model the correlation of data points $t$ at timestamp $x_t$ , consider a local context window of length $K$ as $W_t = \{x_{t-K-1}, ..., x_t\}$, and convert the input data $X$ to a sliding window $W = \{W_1, ..., W_T\}$, using copy-fill for the $t < K$ time points, and append a constant vector $\{x_t, ..., x_t\}$ to the window $W_t$ of length $K - t$ to maintain the size of each window as $K$ .

### 2.3 Data standardization

Since MTS data from the same system are usually multi-source heterogeneous data, which may lead to dimensional differences between different performance metrics, this paper normalizes the datasets used for training and testing to ensure scale consistency of the metrics. The following equations are used for normalization:

$$x_i^{*} = \frac{x_i - x_{\min}}{x_{\max} - x_{\min}} \tag{3}$$

where $x_i^{*}$ denotes the normalized data sample, $x_i$ denotes the original data sample, $x_{\min}$ denotes the minimum value of the training data, and $x_{\max}$ denotes the maximum value of the training data.

## 3. METHODOLOGY

### 3.1 Overview

Figure 1 shows the structure of the designed Trans GAN-WT model. The Trans GAN-WT model as a whole consists of two encoders and decoders. Among them, the encoder converts the time-series data sequence into a low-dimensional coded representation utilizing a self-attention mechanism and a feedforward neural network. The window encoder utilizes the coded representation of the entire sequence generated by the encoder in combination with a sliding window to generate the coded representation of the input window. Subsequently, the decoder inversely decodes the coded representation into a reconstructed sequence that is as similar as possible to the input data through an inverse operation, after which it determines whether an anomaly exists based on the anomaly score of the reconstructed vector. The functions and principles of each module in the model are described in detail next.

### 3.2 Model implementation

#### 3.2.1 Input Module

Since traditional sequential neural networks are inefficient in processing long sequences, the model performance decreases significantly with the increase in sequence length. To alleviate this problem, we use the sliding window technique to partition the long sequence into several shorter windows and batch process the data within each window. The copy-and-fill method is selected to expand the training input $X$ of length $T$ into a sliding window sequence $W=\{W_1,\ldots,W_T\}$. The model uses the window sequence $W$ as the training data and the test input $\hat{X}$ to construct the window sequence $\hat{W}$ for testing.

Considering the continuous nature of the wind turbine dataset, we retained the positional embedding layer of the Transformer. Each time step is assigned a positional vector through this layer. The specific location information embedding method is shown below:

$$PE_{(pos,2i)} = \sin\left(\frac{pos}{10000^{\frac{2i}{d}}}\right)(2i \le d) \tag{4}$$

$$PE_{(pos,2i+1)} = \cos\left(\frac{pos}{10000^{\frac{2i+1}{d}}}\right)(2i \le d) \tag{5}$$

where $PE$ denotes the value of the $i-th$ dimension of the position encoding vector, $pos$ denotes the position index, $i$ denotes the dimension index, and $d$ denotes the dimension of the position encoding vector.

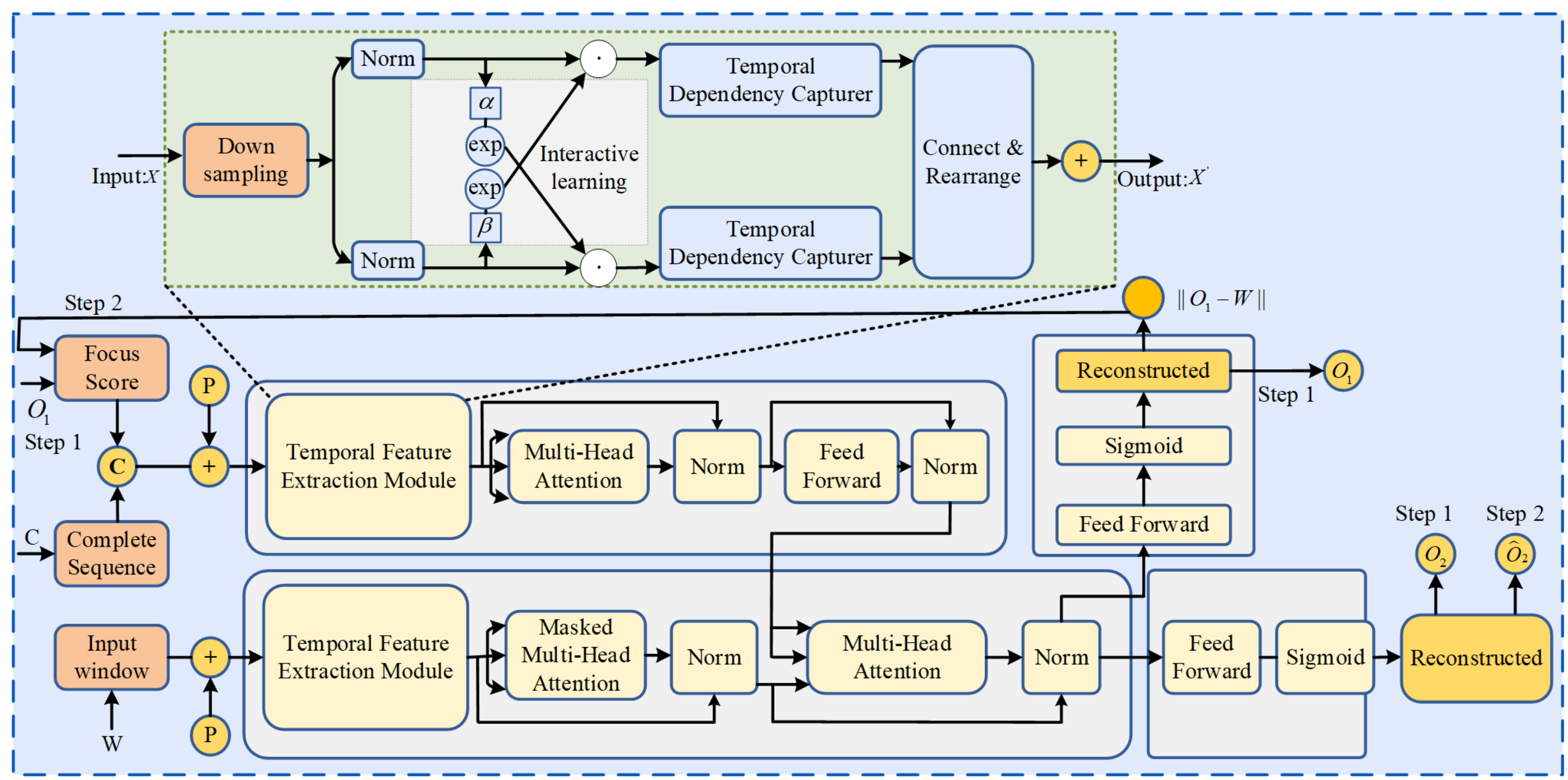

**Figure 1: The overall structure of TransGAN-WT.**

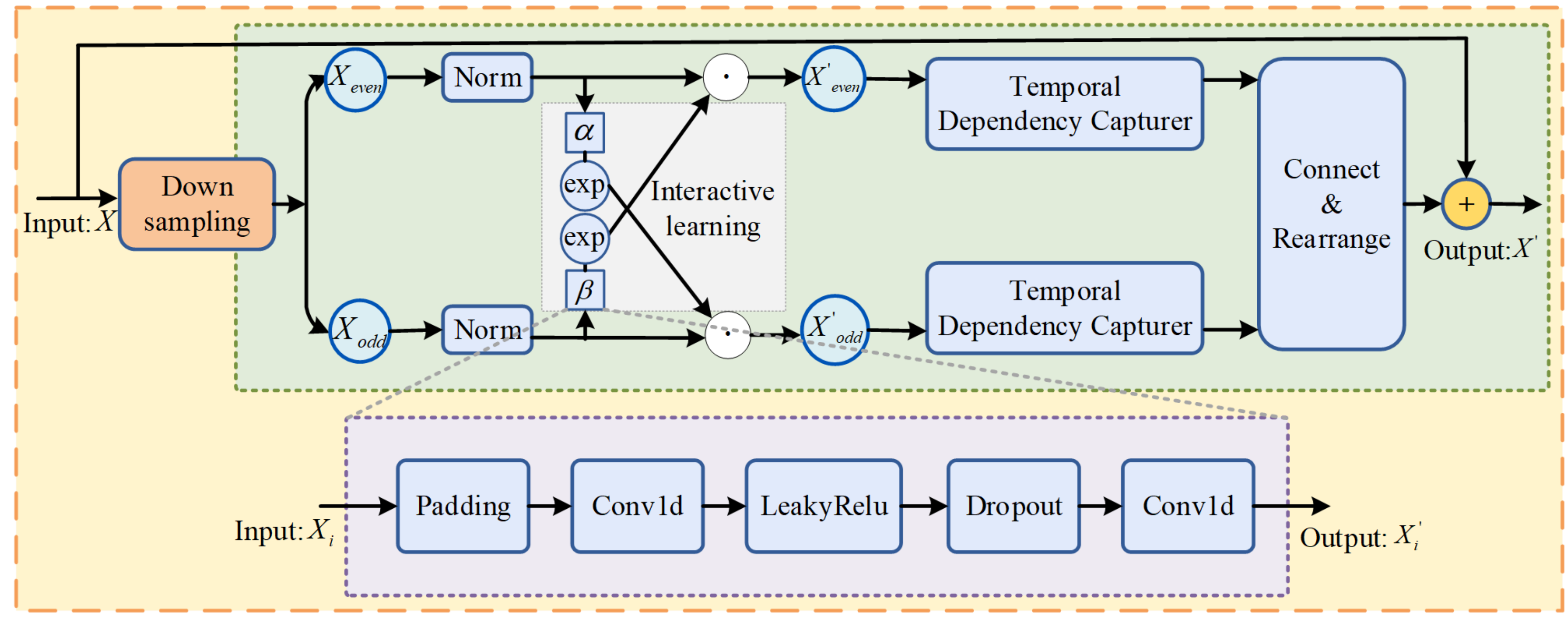

**Figure 2: The detailed structure of the temporal feature extraction module.**

### 3.2.2 Temporal Feature Extraction Module

Based on previous experience, it has been observed that after the wind turbine dataset is downsampled into two sub-sequences, the original temporal dependencies can still be largely preserved. For example, variables such as rotor speed often exhibit trend-like variations over long time scales, while containing fluctuation details over short time scales.

To fully utilize this property, a down-sampling-convolution-interaction architecture is designed in Figure 2, where temporal features are extracted from the down-sampled subsequence by sample convolution and interaction. The down-sampling technique reduces the sampling frequency of the time-series data and captures trends and patterns over longer time scales. The convolution operation can extract features within a local window, making it more suitable for capturing rapid fluctuations under short-term operating conditions, such as power oscillations caused by sudden wind gusts or instantaneous increases in vibration peaks. On the other hand, the interaction operation realizes information exchange by iterating repeatedly at different time resolutions, enabling features at various time scales to influence each other and learn collaboratively, thus enhancing the model's explanatory power.

Permutation Entropy (PE) [30] measures the complexity of time series data. It evaluates the complexity by arranging the time series data and then computing the probability distribution of different permutations, reflecting the data's non-randomness, patterns, and structure. The calculation method is shown in Equation (6):

$$H = -\sum_{i=1}^{T} p_i \log_2(p_i) \tag{6}$$

Where $p_i$ represents the probability of the occurrence of the $i-th$ arrangement among all possible arrangements in the time series data, and $T$ represents the length of the time series data. The smaller the value of arrangement entropy, the more regular the timing data is, and the larger the value, the more complex the timing data is. An adequate representation with enhanced predictive capability can be obtained by recursively extracting and exchanging information at different temporal resolutions. The network structure of the temporal feature extraction module and the relatively low alignment entropy validate the model's potential in handling temporal data with complex temporal dynamics.

As shown in Figure 2, the down-sampling process down samples the original sequence $X$ into two subsequences $X_{odd}$ $X_{even}$ by separating odd and even elements. Although the temporal resolution of these two subsequences is coarser, the majority of the information of the original sequence is still retained. Features are extracted from $X_{odd}$ and $X_{even}$ by using different convolution kernels, respectively.

To compensate for the information loss caused by the down-sampling process,

we design an interactive learning strategy to enhance the feature extraction capability of the model by learning the parameters of each other's affine transformations so that the two subsequences can exchange information. As shown in Figure 2, $X_{odd}$ and $X_{even}$ are projected into the hidden state with two different 1D convolutional modules, $\alpha$ and $\beta$, respectively, then converted to the exponential format and interacted with the elemental level products of $X_{odd}$ and $X_{even}$, as shown in Eqs. (7) and (8).

$$\boldsymbol{X}_{odd}^{'} = \boldsymbol{X}_{odd} \odot \exp(\alpha(\boldsymbol{X}_{even})) \tag{7}$$

$$\boldsymbol{X}_{even}^{'} = \boldsymbol{X}_{even} \odot \exp(\beta(\boldsymbol{X}_{odd})) \tag{8}$$

Which $\alpha$、$\beta$ share the same network architecture, thus ensuring the coordination and consistency among the modules.

Due to variations in sampling frequency under different operating conditions of wind turbines, multi-source SCADA data and condition monitoring data often exhibit heterogeneous data formats. For instance, physical variables such as wind speed, blade pitch angle, yaw, and generator vibration are typically recorded with non-uniform sampling periods. Moreover, owing to factors such as sensor accuracy, environmental noise, and measurement point configuration, the raw time-series data often contain a large amount of redundant or highly correlated information. Such redundancy not only increases the computational burden of subsequent processing but may also mask critical fault signatures.

Against this background, the Multilayer Perceptron (MLP), as a type of fully connected neural network, possesses strong nonlinear mapping and feature extraction capabilities. The MLP can effectively filter out redundant information introduced by differences in sampling mechanisms, environmental disturbances, or duplicated measurement points, while retaining key signals that are closely related to the operational health of the wind turbine. This enables the extraction of feature representations with greater physical relevance and diagnostic value, thereby providing more reliable inputs for subsequent wind turbine fault detection.

To reduce the temporal redundancy in the timing data, the temporal dependency catchers are applied to the down sampled-convolution-interaction timing data $X_{odd}^{'}$ and $X_{even}^{'}$, respectively, to learn the temporal information of these subsequences to capture the dependencies with less redundancy. Figure 3 shows the detailed design of each temporal dependency capturer, consisting of two fully connected layers interspersed with a GELU nonlinear activation function and a Dropout layer. In addition, residual connections are employed to aggregate the input data to the output data, thus enhancing the learning capability of the model. The specific calculation process is shown in Equation (9):

$$\hat{X}^{'} = X^{'} + \left(Dropout\left(GELU\left(X^{'}W_{1}^{T} + b_{1}\right)\right)W_{2}^{T} + b_{2}\right) \tag{9}$$

where $W_1$, $W_2$ denotes the transformation matrix, $b_1$ and $b_2$ indicate the bias quantities.

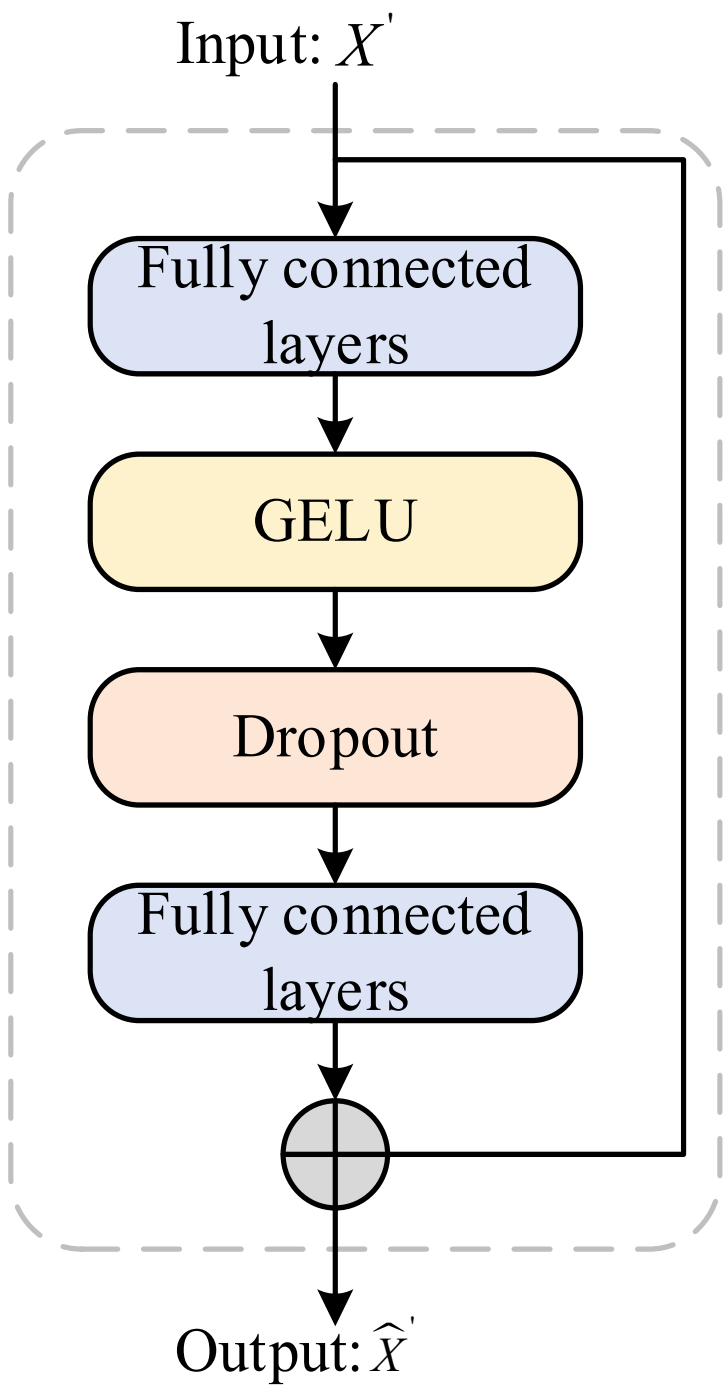

**Figure 3: The diagram of temporal dependency capturer.**

The temporal feature extraction module aggregates the basic information extracted from the two down-sampled subsequences, each providing a local and global view of the entire temporal data. By using a parity reversal splitting operation, the feature elements in all the subsequences are reorganized and concatenated into a new sequence representation. This new sequence representation is then added to the original timing data using residual concatenation to generate a new sequence with enhanced predictability. Finally, this sequence is used as the input to the encoder $S_1$ and the input to the window encoder $S_2$, respectively, and is input to the codec module for computation.

3.2.3 Design of the Encoder–Decoder Module

A multi-head attention mechanism is introduced in the coding and decoding process to capture the relevant dependencies in the timing data. For the inputs $Q$, $K$ and $V$, the output vectors are computed as shown in Equation (10).

$$Attention(Q,K,V) = softmax(\frac{QK^T}{\sqrt{m}})V \tag{10}$$

The multi-head attention mechanism [31] enhances the model's ability to capture diverse information by splitting the input sequence into multiple heads, allowing each head to perform attention computations independently and then aggregating the computation results from these heads. The specific calculation is shown in Equation (11):

$$\begin{aligned} MultiHeadAtt(Q,K,V) = Concat(H_1,...,H_h), \\ where\ H_i = Attention(Q_i,K_i,V_i) \end{aligned} \tag{11}$$

The wind turbine timing data processed through the encoder can generate attentional weights that effectively capture the temporal trends within the input sequence, avoiding the dependence of traditional neural networks on previous timestamped outputs for prediction. Its computational process can be formalized as:

$$S_1^1 = LayerNorm(S_1 + MultiHeadAtt(S_1,S_1,S_1)) \tag{12}$$

$$S_1^2 = LayerNorm(S_1^1 + FeedForward(S_1^1)) \tag{13}$$

In addition, to ensure data consistency during training, a masked multi-attention mechanism is used in the window encoder, as shown in Equation (13), which makes it impossible for the model to use information after the current time step. In this way, the model can only use the information before the current time step to predict the representation of the current time step, thus realizing local contextual attention. The encoding of the entire sequence $S_1^2$ is then used by the window encoder as V and $K$, and the encoded input window is used as the $Q$ matrix for the attention operation.

$$S_2^1 = Mask(MultiHeadAtt(S_2,S_2,S_2)) \tag{14}$$

$$S_2^2 = LayerNorm(S_2 + S_2^1) \tag{15}$$

$$S_2^3 = LayerNorm(S_2^2 + MultiHeadAtt(S_1^2,S_1^2,S_2^2)) \tag{16}$$

The final model is transformed into two output vectors $O_1$ and $O_2$ by two matrices of input $C$ and $W$. The exact calculation is shown in Equation (17):

$$O_i = Sigmoid(FeedForward(S_2^3)) \tag{17}$$

3.2.4 Autoregressive inference module

To address the problem that traditional simple encoder-decoder network structures based on Transformers cannot effectively detect anomalies when the deviation between anomalous wind turbine data and normal data is slight, an autoregressive inference strategy is adopted to achieve significant amplification of anomalous features, with the training process being divided into two stages.

In the first stage, the encoder utilizes a context-based attention mechanism to convert the input window $W$ into a compressed latent representation $S_2^3$. Then, the reconstruction vectors $O_1$ and $O_2$ are computed via Eq. (15). The model generates an approximate reconstruction of the input window employing focus scores. In this process, subsequences with higher deviations are given higher computational weights.

In the second stage, the model retrains the model using the reconstruction loss of the first decoder as the focus score and obtains the output of the second decoder $\hat{O}_2$. By directly feeding back the reconstruction errors generated in the first stage to the attention allocation layer, the model can focus more attention on subsequences that indeed exhibit deviations and may contain anomalies. In this way, anomalous regions can trigger stronger neural network responses, further enhancing the distinction between anomalous and standard features in the model's internal representations.

This two-stage autoregressive inference approach significantly amplifies anomalous features by using reconstruction errors as activations for the encoder's attention mechanism, thereby simplifying the subsequent fault labeling process. Meanwhile, utilizing a windowed encoder to capture short-term temporal trends effectively reduces the probability of false alarms.

### 3.2.5 Adversarial Training and Inference Module

Two independent decoders are used for adversarial training to improve further the robustness and generalization of module training [32]. In the initial stage, the two decoders independently reconstruct the input window. The reconstruction process of each decoder in the first stage is shown in Eqs. (18) and (19):

$$L_1 = || O_1 - W ||_2 \tag{18}$$

$$L_2 = || O_2 - W ||_2 \tag{19}$$

As shown in Equation (18), the second decoder goal aims to distinguish the input window from the reconstructed sequence by maximizing the difference of $|| \hat{O}_2 - W ||_2$. On the other hand, the first decoder defrauds the second decoder by perfectly reconstructing the input window, which induces the second decoder to generate the same output as $O_2$ at this stage, thus achieving a match with the input in stage 1. This adversarial interaction is formally captured by the following minimax optimization over their respective parameters:

$$\theta_1^* = \arg\min_{\theta_1} \max_{\theta_2} || \hat{\boldsymbol{O}}_2(\theta_2; \theta_1) - \boldsymbol{W} ||_2 \tag{20}$$

where $\theta_1$ and $\theta_2$ denote the trainable parameters of the first and second decoders, respectively. Thereby, the goal of the first decoder is to minimize the reconstruction error of the self-regulated outputs, while the goal of the second decoder is to maximize the reconstruction error of the self-regulated outputs, which we achieve through adversarial loss, as shown in Eqs. (21) and (22):

$$L_1 = + || \hat{O}_2 - W ||_2 \tag{21}$$

$$L_2 = - || \hat{O}_2 - W ||_2 \tag{22}$$

In turn, the evolutionary loss function that combines the reconstruction loss and the adversarial loss at both stages is shown in Eqs. (23) and (24):

$$L_1 = \varepsilon^{-n} \| O_1 - W \|_2 + (1 - \varepsilon^{-n}) \| \hat{O}_2 - W \|_2 \tag{23}$$

$$L_2 = \varepsilon^{-n} \| O_2 - W \|_2 - (1 - \varepsilon^{-n}) \| \hat{O}_2 - W \|_2 \tag{24}$$

where $\varepsilon$ denotes the training parameter close to 1 and $n$ denotes the number of iterations.

## 4. Experiments and analysis

### 4.1 Experimental preparation

Data selection. The dataset used in this paper comes from the operation records of wind turbines in a wind farm in China. These data are obtained directly from the daily work production of wind turbines and reflect the actual performance in the natural working environment. Influenced by realistic environmental factors, each wind turbine's raw data includes many abnormal data points. These data match the actual application requirements of the real-time monitoring and control system (SCADA) for wind turbines regarding abnormal data detection. The dataset contains 12 groups of wind turbine data, totaling 497,837 entries. Among them, the data volume of Unit 5 is the largest, with 50,962 data items, and the data volume of Unit 10 is the smallest, with 30,184 data items, and the details are shown in Table 2

**Table 2: Data volume statistics of wind turbine dataset.**

| Dataset groups | Data volume |
|---|---|
| 1# | 40727 |
| 2# | 38855 |
| 3# | 38995 |
| 4# | 44335 |
| 5# | 50962 |
| 6# | 45592 |
| 7# | 43324 |
| 8# | 38470 |
| 9# | 42824 |
| 10# | 30184 |
| 11# | 36848 |
| 12# | 46721 |
| Total number of datasets | 497837 |

Based on the experience and knowledge of past experts and the original dataset, we filtered out 11 measurements related to changes in generator performance, as shown in Table 3.

**Table 3: Detailed attribute information of the dataset**

| No. | Features |
|---|---|
| 1 | Timestamp |
| 2 | Wind Speed |
| 3 | Power |
| 4 | Wind direction |
| 5 | Wheel Diameter |
| 6 | Fan No. |
| 7 | Rated power |
| 8 | Cut-in wind speed |
| 9 | Cutting out wind speed |
| 10 | Rotor speed |
| 11 | Wind wheel speed range |

The wind turbine dataset used in this study contains 11 feature indicators. Since some of them (e.g., timestamp, rotor diameter, turbine ID, and rated power) are fixed or categorical, our analysis focuses on the continuously varying features, namely wind speed, power, wind direction, rotor speed, cut-in wind speed, and cut-out wind speed. To comprehensively describe the distribution of these variables, we computed their Maximum (Max), Minimum (Min), Average (Avg), Median (Med), Standard Deviation (SD), First Quartile (Q1), Third Quartile (Q3), Skewness (Skew), and Kurtosis (Kurt). These statistical results provide essential support for subsequent anomaly detection modeling, as summarized in Table 4.

**Table 4: Statistical analysis of features**

| No. | Max | Min | Avg | Med | SD | Q1 | Q3 | Skew | Kurt |
|---|---|---|---|---|---|---|---|---|---|
| Wind speed | 16.2 | 0 | 4.82 | 4.7 | 2.32 | 3 | 6.4 | 0.39 | |
| Power | 96700 | 9516.51 | 28363.67 | 19018.8 | 21688.43 | 9848.54 | 42745.12 | 0.96 | |
| Wind direction | 359 | 0 | 134.08 | 104 | 92 | 71 | 190.25 | 0.9 | |
| Rotor speed | 16.8 | 0 | 7.64 | 8.01 | 3.5 | 5.5 | 10.8 | 0.27 | |
| Cut-in wind speed | 3.7 | 3 | 3.28 | 3.11 | 0.18 | 3.05 | 3.32 | 0.76 | |
| Cut-out wind speed | 26.6 | 25 | 25.9 | 25.96 | 0.44 | 25.55 | 26.37 | 0.21 | |

Evaluation indicators: Given the severe imbalance in the dataset for wind turbine multivariate time-series anomaly detection, where the number of anomalous samples is much smaller than that of standard samples, three metrics based on the confusion matrix are typically used to evaluate model performance. Precision (P), Recall rate (R), False positive rate (FPR), and F1 score (F1). Precision (P) measures the proportion of all results predicted to be positive samples that are positive samples. Recall (R) reflects the proportion of samples that are correctly predicted as positive among those that are positive. The F1 score (F1) is the reconciled average of Precision (P) and Recall (R), and the higher its value, the more balanced and stable the model's performance is. The specific formula is as follows:

$$P = \frac{TP}{TP + FP} \tag{25}$$

$$R = \frac{TP}{TP + FN} \tag{26}$$

$$FPR = \frac{FP}{FP + TN} \tag{27}$$

$$F1 = 2 \times \frac{P \times R}{P + R} \tag{28}$$

where True Positive ($TP$) denotes samples with both actual and predicted normal, False Negative ($FN$) denotes samples with true normal but predicted abnormal, False Positive ($FP$) denotes samples with actual abnormal but predicted normal, and True Negative ($TN$) indicates a sample that is abnormal in both truth and prediction.

Device parameters. The experimental design of the proposed algorithm and baseline model was carried out under the hardware configuration of a 64-bit Windows server equipped with Intel Core i9-10900X CPU, GeForce RTX 2080 GPU, 1TB hard disk, and 32G RAM. The operating system version is Linux CentOS 7, the Python version is 3.8, and the integrated development environment is VSCode. Considering the actual operational and maintenance environment of wind farms, the recommended minimum hardware configuration includes a GPU with 8GB memory (such as GTX1070 or equivalent performance graphics card), a 4-core CPU, and 16GB system memory, which can meet the real-time anomaly monitoring requirements for single or a small number of wind turbines. To ensure training efficiency and stability of parallel monitoring for multiple turbines, the recommended configuration includes a GPU with 12GB or more memory and an 8-core or higher CPU.

Experimental parameter settings. We optimize the algorithm model using the Adam optimizer that adaptively adjusts the learning rate, presetting the initial learning rate to 0.01 and reducing the learning rate to half of the original on each training round. The model-related parameter settings are shown in Table 5.

**Table 5 Experiment-related parameter settings**

| Parameter Name | Parameter values |
|---|---|
| Batch size | 128 |
| Sliding window size | 30 |
| Number of encoder layers | 1 |
| Number of feedforward unit layers | 2 |
| Number of training rounds | 30 |
| Learning rate | 0.01 |

| Dropout | 0.1 |
|---|---|

### 4.2 Performance comparison with other models

To evaluate the effectiveness of our proposed Trans GAN-WT, the following five baseline models are selected for comparison: SVR [33], LSTM-NDT [34], MAD-GAN [35], MTAD-GAT [36], GDN [37], SOEP-KNN[38], and Tran AD [39]. The models are described as follows:

- SVR: The model adopts a combination of a grey correlation-based algorithm to construct feature parameters and support vector regression (SVR) for wind turbine condition monitoring, and uses a genetic algorithm to optimize the model parameters and achieve anomaly detection through distance threshold analysis.
- LSTM-NDT: The LSTM model is used as the basis for prediction. The prediction error identifies anomalies, and a parameter-free dynamic threshold estimation method is introduced to realize anomaly detection in multidimensional data.
- MAD-GAN: The MAD-GAN model uses a generative adversarial network, with the generator creating synthetic time-series data and the discriminator distinguishing between accurate and generated data. The discriminator and generator are also based on an extended short-term memory network.
- MTAD-GAT: The MTAD-GAT model learns the complex dependencies of multivariate time-series data in time and feature dimensions using two parallel graph attention layers and jointly optimizing the prediction-based and reconstruction-based models combined with individual timestamp predictions to obtain a better temporal representation.
- GDN: The GDN model employs a graph neural network based on an attention mechanism to explicitly learn relational dependency graphs between variables and feeds this learned relational graph information into a prediction network to identify anomalies.
- SOEP-KNN: This study proposes a cost-sensitive, real-time anomaly detection framework using wind power and wind speed features from Irish wind farm SCADA data, enabling early and accurate detection of wind turbine downtime events without strong distributional assumptions.
- Tran AD: This is a Transformer-based anomaly detection model. It learns data representation by encoding the time-series data and uses the reconstruction error to identify anomalies. A Model-Agnostic Meta-Learning (MAML) strategy is used in the model training process, which helps the model to quickly learn the temporal trends of the input data on a limited dataset.

(1) To ensure the objectivity of the model evaluation, we strictly followed the parameter settings used in the original papers of each model. We actively searched for the optimal hyperparameters using a grid search method. In addition, each model was tested by seven repetitions, using different random seeds and data division strategies, and obtaining the average value of the evaluation indexes. The specific experimental results are shown in Table 6.

**Table 6 Detection results of anomalous data of wind turbines with different models. We will present the relevant results in the following four sub-tables**

(a) Detection results of anomaly data from different models for wind turbines 1-3.

| Method | 1# | | | | 2# | | | | 3# | | | |
|---|---|---|---|---|---|---|---|---|---|---|---|---|
| | *P* (%) | *R* (%) | *FPR(%)* | *F*1 (%) | *P* (%) | *R* (%) | *FPR(%)* | *F*1 (%) | *P* (%) | *R* (%) | *FPR(%)* | *F*1 (%) |
| SVR | 77.32 | 72.14 | 0.44 | 73.44 | 77.45 | 70.82 | 0.53 | 73.95 | 72.78 | 69.36 | 0.64 | 71.02 |
| LSTM-NDT | 82.97 | 71.41 | 0.25 | 76.76 | 83.23 | 72.67 | 0.28 | 77.59 | 81.73 | 78.12 | 0.28 | 79.88 |
| MAD-GAN | 81.85 | 85.53 | 0.13 | 83.65 | 88.71 | 76.92 | 0.19 | 82.40 | 82.21 | 89.23 | 0.17 | 85.58 |
| MTAD-GAT | 81.40 | 92.21 | 0.09 | 86.47 | 89.97 | 91.06 | 0.10 | 90.51 | 88.43 | 82.31 | 0.21 | 85.26 |
| GDN | 77.51 | 94.64 | 0.11 | 85.22 | 90.25 | 89.14 | 0.12 | 89.69 | 89.45 | 81.18 | 0.15 | 85.11 |
| SOEP-KNN | 91.89 | 93.75 | 0.08 | 90.17 | 94.71 | 92.55 | 0.11 | 89.59 | 90.07 | 91.89 | 0.18 | 89.62 |
| Tran AD | 90.57 | 95.29 | 0.09 | 92.87 | 88.21 | 95.74 | 0.08 | 91.82 | 90.47 | 87.43 | 0.13 | 88.92 |
| Trans GAN-WT | **94.23** | **98.20** | **0.04** | **96.17** | **97.80** | **95.32** | **0.03** | **96.54** | **96.71** | **94.27** | **0.08** | **95.47** |

(b) Detection results of anomaly data from different models for wind turbines 4-6.

| Method | 4# | | | | 5# | | | | 6# | | | |
|---|---|---|---|---|---|---|---|---|---|---|---|---|
| | *P* (%) | *R* (%) | *FPR(%)* | *F*1 (%) | *P* (%) | *R* (%) | *FPR(%)* | *F*1 (%) | *P* (%) | *R* (%) | *FPR(%)* | *F*1 (%) |
| SVR | 71.09 | 73.47 | 0.57 | 72.66 | 71.50 | 74.62 | 0.49 | 73.53 | 70.14 | 71.52 | 0.47 | 71.08 |
| LSTM-NDT | 72.81 | 76.29 | 0.36 | 74.51 | 78.49 | 70.67 | 0.29 | 74.38 | 79.53 | 78.12 | 0.29 | 78.82 |
| MAD-GAN | 92.32 | 86.59 | 0.27 | 89.36 | 84.47 | 80.18 | 0.19 | 82.27 | 74.21 | 79.23 | 0.33 | 76.64 |
| MTAD-GAT | 84.42 | 89.72 | 0.38 | 86.99 | 89.67 | 88.06 | 0.27 | 88.86 | 73.43 | 80.31 | 0.39 | 76.72 |
| GDN | 91.51 | 88.12 | 0.17 | 89.78 | 91.46 | 92.14 | 0.16 | 91.80 | 89.45 | 84.18 | 0.27 | 86.74 |
| SOEP-KNN | 92.74 | 91.58 | 0.19 | 88.86 | 93.74 | 93.07 | 0.21 | 87.48 | 86.71 | 84.71 | 0.26 | 84.11 |
| Tran AD | 92.69 | 94.69 | 0.11 | 93.68 | 93.21 | 92.43 | 0.21 | 92.82 | 85.47 | 91.43 | 0.18 | 88.35 |
| Trans GAN-WT | **94.59** | **97.92** | **0.06** | **96.23** | **98.83** | **97.19** | **0.05** | **98.00** | **94.71** | **97.27** | **0.09** | **95.97** |

(c) Detection results of anomaly data from different models for wind turbines 7-9.

| Method | 7# | | | | 8# | | | | 9# | | | |
|---|---|---|---|---|---|---|---|---|---|---|---|---|
| | *P* (%) | *R* (%) | *FPR*(%) | *F*1 (%) | *P* (%) | *R* (%) | *FPR*(%) | *F*1 (%) | *P* (%) | *R* (%) | *FPR*(%) | *F*1 (%) |
| SVR | 72.91 | 74.59 | 0.58 | 73.96 | 74.81 | 69.08 | 0.41 | 72.93 | 70.65 | 72.83 | 0.48 | 71.82 |
| LSTM-NDT | 83.18 | 79.32 | 0.16 | 81.20 | 79.23 | 71.67 | 0.26 | 75.26 | 71.67 | 74.12 | 0.33 | 72.87 |
| MAD-GAN | 89.51 | 86.43 | 0.17 | 87.94 | 83.42 | 87.13 | 0.19 | 85.23 | 82.39 | 86.84 | 0.41 | 84.56 |
| MTAD-GAT | 82.44 | 91.43 | 0.07 | 86.70 | 89.67 | 91.06 | 0.16 | 90.36 | 87.43 | 82.31 | 0.26 | 84.79 |
| GDN | 85.51 | 91.29 | 0.11 | 88.31 | 90.46 | 92.14 | 0.09 | 91.29 | 89.45 | 91.18 | 0.17 | 90.31 |
| SOEP-KNN | 90.71 | 93.06 | 0.16 | 92.17 | 93.58 | 94.77 | 0.15 | 93.89 | 91.07 | 95.38 | 0.17 | 93.74 |
| Tran AD | 91.42 | 94.71 | 0.09 | 93.04 | 97.21 | 94.43 | 0.12 | 95.80 | 92.47 | 90.43 | 0.19 | 91.44 |
| Trans GAN-WT | **94.28** | **97.29** | **0.08** | **95.74** | **97.8** | **96.32** | **0.08** | **97.05** | **97.71** | **98.27** | **0.12** | **97.99** |

(d) Detection results of anomaly data from different models for wind turbines 10-12.

| Method | 10# | | | | 11# | | | | 12# | | | |
|---|---|---|---|---|---|---|---|---|---|---|---|---|
| | *P* (%) | *R* (%) | *FPR*(%) | *F*1 (%) | *P* (%) | *R* (%) | *FPR*(%) | *F*1 (%) | *P* (%) | *R* (%) | *FPR*(%) | *F*1 (%) |
| SVR | 70.69 | 72.48 | 0.42 | 71.50 | 71.58 | 73.91 | 0.38 | 72.55 | 72.68 | 71.06 | 0.46 | 71.92 |
| LSTM-NDT | 74.36 | 78.32 | 0.25 | 76.29 | 72.23 | 75.67 | 0.16 | 73.91 | 77.61 | 72.69 | 0.17 | 75.07 |
| MAD-GAN | 79.32 | 81.53 | 0.18 | 80.41 | 81.42 | 83.13 | 0.20 | 82.27 | 81.37 | 84.23 | 0.15 | 82.78 |
| MTAD-GAT | 84.42 | 87.11 | 0.18 | 85.74 | 84.58 | 83.16 | 0.19 | 83.86 | 84.37 | 82.91 | 0.27 | 83.63 |
| GDN | 88.51 | 85.12 | 0.14 | 86.78 | 88.27 | 81.74 | 0.14 | 84.88 | 89.36 | 84.29 | 0.33 | 86.75 |
| SOEP-KNN | 90.05 | 93.28 | 0.19 | 91.77 | 92.71 | 92.71 | 0.18 | 91.64 | 90.75 | 87.39 | 0.19 | 89.91 |
| Tran AD | 91.42 | 94.17 | 0.11 | 92.77 | 93.21 | 90.43 | 0.11 | 91.80 | 92.47 | 91.43 | 0.15 | 91.95 |
| Trans GAN-WT | **96.23** | **95.88** | **0.07** | **96.05** | **97.8** | **96.32** | **0.06** | **97.05** | **96.71** | **93.27** | **0.11** | **94.96** |

The experimental results reveal that Trans GAN-WT obtains different degrees of improvement in each evaluation index on the dataset of 12 different wind turbines compared with other baseline models. However, the number of anomalies in the various wind turbine datasets varies due to abnormal events such as extreme weather and unplanned downtime. The sparse positive samples and large feature dimensions make the multivariate wind turbine timing data relationship more challenging to mine. Compared with other groups of unit datasets, the anomaly detection accuracy of 1# and 7# units is significantly lower than that of different groups. In addition, it can be seen that the FPR values of all models are relatively low (0.07% to 0.41%), indicating that the false positive rate of these models is generally controllable during anomaly detection.

From the overall results, the Trans GAN-WT model improves the F1 value by 7.64% compared with the GDN model, which verifies that capturing long-term dependencies in the data through the self-attention mechanism is better than capturing temporal dependencies through the relationship graph information. The Trans GAN-WT model improves the F1 value by 2.35% compared with the Tran AD model, and the Trans GAN-WT model improves the F1 value by 2.35% compared with the Tran AD model. The Trans GAN-WT model shows a more vital ability to capture anomalies. Compared with the MTAD-GAT model, the Trans GAN-WT model shows a more significant improvement, with a 6.17% increase in the F1 value. The SOEP-KNN method relies on statistical domain features for pattern matching between similar operating conditions of wind turbines, demonstrating certain competitive performance. However, TransGAN-WT still exhibits significant advantages. Although SOEP-KNN can benefit from local similarity information—for instance, by comparing the current operating state with the nearest neighbor data under similar wind speed and load conditions to detect anomalies—TransGAN-WT has a stronger advantage in capturing long-range dependencies and global temporal correlations of SCADA signals. This enables it to more robustly identify subtle downtime symptoms and damage evolution processes in complex wind turbine operations.

The decrease in the MTAD-GAT model's performance is mainly because it assumes that the relationship between the variables is a complete graph in acquiring spatial information, which contradicts the incomplete directed graphs in natural industrial environments. Compared with the LSTM-NDT and MAD-GAN models, the F1 value of the Trans GAN-WT model is improved by 16.88% and 7.38%, respectively, indicating that it is better at capturing anomaly information. These two baseline models are deficient in capturing anomaly information due to their neglect of learning the low-dimensional representation of each metric along the time dimension, thus leading to their deficiencies in capturing anomaly information. Therefore, considering the collaborative learning of features at different time scales and eliminating temporal redundancy is crucial for anomaly data detection.

(2) To assess the statistical significance of the performance differences between Trans GAN-WT and the other baseline models, we performed K-fold cross-validation (K=10). We recorded the F1 score for each fold. We analyzed the results using the nonparametric Wilcoxon signed-rank test, and the test results are shown in Table 7.

**Table 7 Comparison of Wilcoxon signed-rank test results**

| Method | Average F1 value | Standard deviation | vs Trans GAN-WT | | |
|---|---|---|---|---|---|
| | | | Z-statistic | p-value | significance |
| TransGAN-WT | 0.950 | 0.008 | - | - | - |
| SVR | 0.864 | 0.012 | -2.803 | p < 0.01 | *** |
| LSTM-NDT | 0.886 | 0.011 | -2.701 | p < 0.01 | *** |
| MAD-GAN | 0.907 | 0.010 | -2.599 | p < 0.01 | *** |
| MTAD-GAT | 0.922 | 0.009 | -2.497 | p < 0.05 | ** |
| SOEP-KNN | 0.928 | 0.009 | -2.418 | p < 0.01 | *** |
| GDN | 0.918 | 0.013 | -2.548 | p < 0.05 | ** |

Where: ***: $p < 0.01$ (highly significant), **: $p < 0.05$ (significant), *: $p < 0.1$ (marginally significant)

As observed in Table 5, all p-values are less than 0.05, indicating that the performance improvement of TransGAN-WT is statistically significant and not due to random fluctuations. In particular, the improvements over traditional methods (SVR, LSTM-NDT, MAD-GAN) reach a highly significant level ($p < 0.01$), demonstrating the effectiveness and reliability of the proposed method.

(3) Training time and theoretical time complexity analysis. Model size and training time, to some extent, reflect the computational efficiency and complexity of the algorithm. During model training, we conducted a detailed statistical analysis of the key metrics of Trans GAN-WT: the model has 29 million parameters, a storage size of 78 MB, GPU memory usage of 1.28 GB during inference, and 1.93 GB during training. To comprehensively evaluate the computational efficiency of the model, we present in Table 8 the training time comparison results between Trans GAN-WT and various baseline algorithms on 12 wind turbine datasets, thereby providing quantitative analytical evidence for the efficiency-performance trade-off of different algorithms.

Table 8 Comparison with existing wind turbine anomaly detection models based on training time. All training times are reported in seconds (s).

| Dataset groups | Method | | | | | | |
|---|---|---|---|---|---|---|---|
| | SVR | LSTM-NDT | MAD-GAT | MTAD-GAT | GDN | SOEP-KNN | TranAD | TransGAN-WT |
| 1# | 23.0 | 58.2 | 248.5 | 187.4 | 103.9 | 18.1 | 156.6 | 312.4 |
| 2# | 23.3 | 59.0 | 251.6 | 189.9 | 105.2 | 19.0 | 158.7 | 316.5 |
| 3# | 18.3 | 46.3 | 197.6 | 149.0 | 82.7 | 18.9 | 124.6 | 248.6 |
| 4# | 26.1 | 66.1 | 281.9 | 212.5 | 117.9 | 23.8 | 177.7 | 354.5 |
| 5# | 21.2 | 53.7 | 228.9 | 172.6 | 95.8 | 19.2 | 144.4 | 288.0 |
| 6# | 23.9 | 60.5 | 258.2 | 194.8 | 108.0 | 22.4 | 162.8 | 324.8 |
| 7# | 20.5 | 51.9 | 221.4 | 167.0 | 92.7 | 19.2 | 139.7 | 278.8 |
| 8# | 24.2 | 61.3 | 261.0 | 196.9 | 109.3 | 23.8 | 164.8 | 328.9 |
| 9# | 23.4 | 59.3 | 252.0 | 190.2 | 105.6 | 20.1 | 159.2 | 317.5 |
| 10# | 15.6 | 39.5 | 168.4 | 127.0 | 70.5 | 14.3 | 106.2 | 211.9 |
| 11# | 18.7 | 47.4 | 202.0 | 152.4 | 84.5 | 18.9 | 127.4 | 254.3 |
| 12# | 22.9 | 58.0 | 247.3 | 186.5 | 103.5 | 21.9 | 156.0 | 311.2 |

As shown in Table 8, the algorithms exhibit a clear hierarchical pattern in terms of computational efficiency. The SVR algorithm demonstrates the shortest training time (average 21.8s) with a time complexity of $O(n^2)$, representing a typical traditional machine learning approach. However, although SVR has the greatest advantage in runtime, its ability to capture the complex temporal dependencies and multivariate nonlinear relationships of wind turbine systems is limited, and its anomaly detection performance is significantly inferior to that of deep learning methods. The LSTM-NDT algorithm requires an average training time of 55.1s, approximately 2.5 times longer than SVR, with a time complexity of $O(n \cdot d^2)$. It can handle basic temporal dependencies and shows certain sensitivity to typical anomalies such as temperature rise trends or load fluctuations. Nevertheless, when dealing with more complex operational variations—such as power anomalies caused by sudden wind speed changes or gearbox fatigue characteristics—its detection performance remains insufficient. The MTAD-GAT algorithm integrates a graph attention network with a multi-task learning framework, reaching a complexity of $O(T |V|^2 \cdot d)$. Its training time is 8.1 times longer than SVR, reflecting the cumulative effect of multiple complex mechanisms. This higher computational cost provides an advantage in modeling interdependencies among different wind turbine subsystems (e.g., correlations between main shaft, gearbox, and generator vibration signals). However, the adversarial training process involving a generator and multiple discriminators substantially impacts training efficiency. Although the TransGAN-WT algorithm has the highest training time, with an average of 295.6 seconds, which is 13.5 times longer than SVR, this increase in time cost brings significant performance improvements, enabling it to have strong temporal modeling capabilities and the ability to capture global relationships. While it has the highest computational complexity, in safety-critical applications like wind turbine anomaly detection, the trade-off of additional time cost for higher detection accuracy and lower false negative rates is well worth it.

### 4.3 Ablation experiment

To verify the effectiveness of each module of the proposed model, Trans GAN-WT, we will conduct ablation experiments to verify and analyze the role of the critical modules of the model. Three model variants, model-trains, model-conv -conv, and model-map, are designed, and the experimental results are shown in Table 9. Among them:

- model-tran: this variant removes the generative adversarial network and autoregressive inference modules and retains only the Transformer's original structure and the temporal feature extraction module.

- model-conv: this variant removes the down-sampling-convolution-interaction architecture and retains only the time-dependent catcher to reduce the temporal redundancy in the timing data.
- model-mlp: this variant removes the temporal dependency capturer. It retains only the down-sampling-convolution-interaction architecture to aggregate the basic information extracted from the down-sampled subsequences, facilitating the interaction and learning of features from different time scales.

As shown in Table 5, the effectiveness of the components in the Trans GAN-WT model in the anomaly detection process is observed. When the generative adversarial network and autoregressive inference module are removed, the F1 mean value of the model on each data set decreases by 19.19%, verifying the significant effect of the reconstruction error and autoregressive inference by amplifying them on improving the detection. The two-stage autoregressive inference uses reconstruction loss feedback as the basis for attention weight adjustment, enabling the model to alternately optimize the feature space and gradually increase the distance between anomalous and normal samples, significantly improving the separability between small-deviation anomalies and standard samples. Experimental results demonstrate that this strategy can achieve high levels of recall and detection sensitivity for small-magnitude anomalies, highlighting the significant role of amplifying reconstruction errors and autoregressive inference in enhancing detection performance. The F1 mean value of the model on each dataset decreases by 12.50% when the down-sampling-convolution-interaction architecture is removed, indicating that this architecture is crucial for the interaction and learning of features at different time scales, which helps the model obtain more accurate feature representations. When the temporal dependency capturer is removed, the F1 value of the model on each dataset decreases by 0.53% compared with that of the entire model, and the effect varies in different groups of datasets, which indicates that the other degree of temporal redundancy in the dataset affects the effect of the time-dependent catcher, and also verifies the effectiveness of the time-dependent catcher in the anomaly detection process.

**Table 9 Results of ablation experiments with different wind turbine datasets.**

| Data set groups | model-tran | | | model -conv | | | model-mlp | | | original model | | |
|---|---|---|---|---|---|---|---|---|---|---|---|---|
| | $P$ (%) | $R$ (%) | $F1$ (%) | $P$ (%) | $R$ (%) | $F1$(%) | $P$ (%) | $R$ (%) | $F1$ (%) | $P$ (%) | $R$ (%) | $F1$ (%) |
| 1# | 71.89 | 72.99 | 72.44 | 87.49 | 88.91 | 88.19 | 93.47 | 97.61 | 95.50 | **94.23** | **98.20** | **96.17** |
| 2# | 74.68 | 72.98 | 73.82 | 92.01 | 90.81 | 91.41 | 96.91 | 94.61 | 95.75 | **97.80** | **95.32** | **96.54** |
| 3# | 71.24 | 73.26 | 72.24 | 83.27 | 85.94 | 84.58 | 93.87 | 94.01 | 93.94 | **96.71** | **94.27** | **95.47** |
| 4# | 77.35 | 78.54 | 77.94 | 87.25 | 83.59 | 85.38 | 94.55 | 96.94 | 95.73 | **94.59** | **97.92** | **96.23** |
| 5# | 74.98 | 71.23 | 73.06 | 87.69 | 82.97 | 85.26 | 96.58 | 95.81 | 96.19 | **98.83** | **97.19** | **98.00** |
| 6# | 72.49 | 69.98 | 71.21 | 85.44 | 86.92 | 86.17 | 93.82 | 95.49 | 94.65 | **94.71** | **97.27** | **95.97** |
| 7# | 78.41 | 79.47 | 78.94 | 86.59 | 83.29 | 84.91 | 93.75 | 96.88 | 95.29 | **94.23** | **97.29** | **95.74** |
| 8# | 69.94 | 72.69 | 71.29 | 89.75 | 88.01 | 88.87 | 96.92 | 96.01 | 96.46 | **97.8** | **96.32** | **97.05** |
| 9# | 77.49 | 79.67 | 78.56 | 90.26 | 91.64 | 90.94 | 94.97 | 97.59 | 96.26 | **97.71** | **98.27** | **97.99** |
| 10# | 72.67 | 73.77 | 73.22 | 93.89 | 92.97 | 93.43 | 94.86 | 94.99 | 94.92 | **96.23** | **95.88** | **96.05** |
| 11# | 72.88 | 71.92 | 72.40 | 90.68 | 88.64 | 89.65 | 95.94 | 93.21 | 94.56 | **97.8** | **96.32** | **97.05** |
| 12# | 64.96 | 65.31 | 65.13 | 85.89 | 82.66 | 84.24 | 94.86 | 91.77 | 93.29 | **96.71** | **93.27** | **94.96** |

### 4.4 Parametric experiment

The size of the time window can significantly impact several aspects of the model's detection effectiveness and generalization ability. When the window is small, the model is better suited to capture short-term anomalies but is more susceptible to data noise. On the contrary, if the window is too large, short anomalies may be hidden in a large number of data points in such a window, which may lead to a decrease in the F1 value of the model. Figure 4 illustrates the F1 values of Trans GAN-WT with different window sizes.

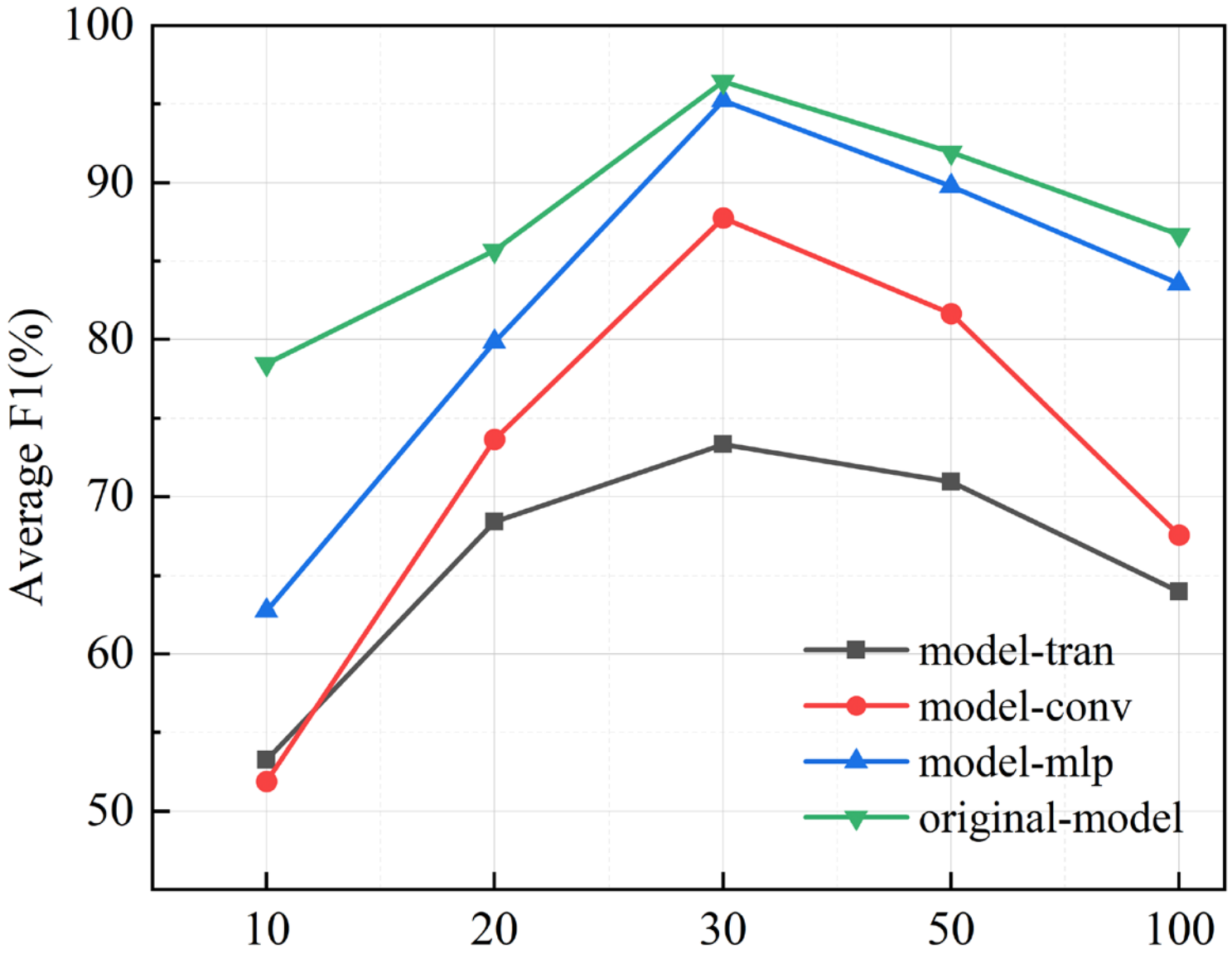

**Figure 4 Sensitivity to the window size.**

Observation of Figure 4 shows that a time window that is too large or too small affects the detection performance, and the best F1 score and suitable detection granularity can be obtained when the window size is 30.

### 4.5 Visualization of results

To further validate the interpretability and practical reliability of the TransGAN-WT model, we employ a visualization-based analysis of the model's reconstructed outputs to illustrate its anomaly detection mechanism. Since TransGAN-WT is trained on SCADA data from wind turbines under diverse temporal sequences, it learns and captures both the structural characteristics and temporal patterns of wind turbine operations. During the testing phase, when the input data shares similar structural features with the training data, the model is able to accurately reconstruct the original signals; conversely, significant reconstruction errors occur during abnormal periods. Given that deep learning-based anomaly detection methods generally exhibit a "black-box" nature, providing only the anomaly score may be insufficient to support rapid decision-making in practical engineering operation and maintenance. Therefore, it is necessary to provide interpretability evidence for the detection results to mitigate economic and safety risks caused by potential misdiagnoses. To demonstrate the practical effectiveness of this mechanism, we conduct a case study using a wind turbine SCADA dataset, analyzing a real fault event caused by severe wear of the generator's rear bearing. This event was discovered during an on-site inspection and subsequently confirmed by experts, after which it was incorporated into the validation set as an anomaly sample with a known root cause.

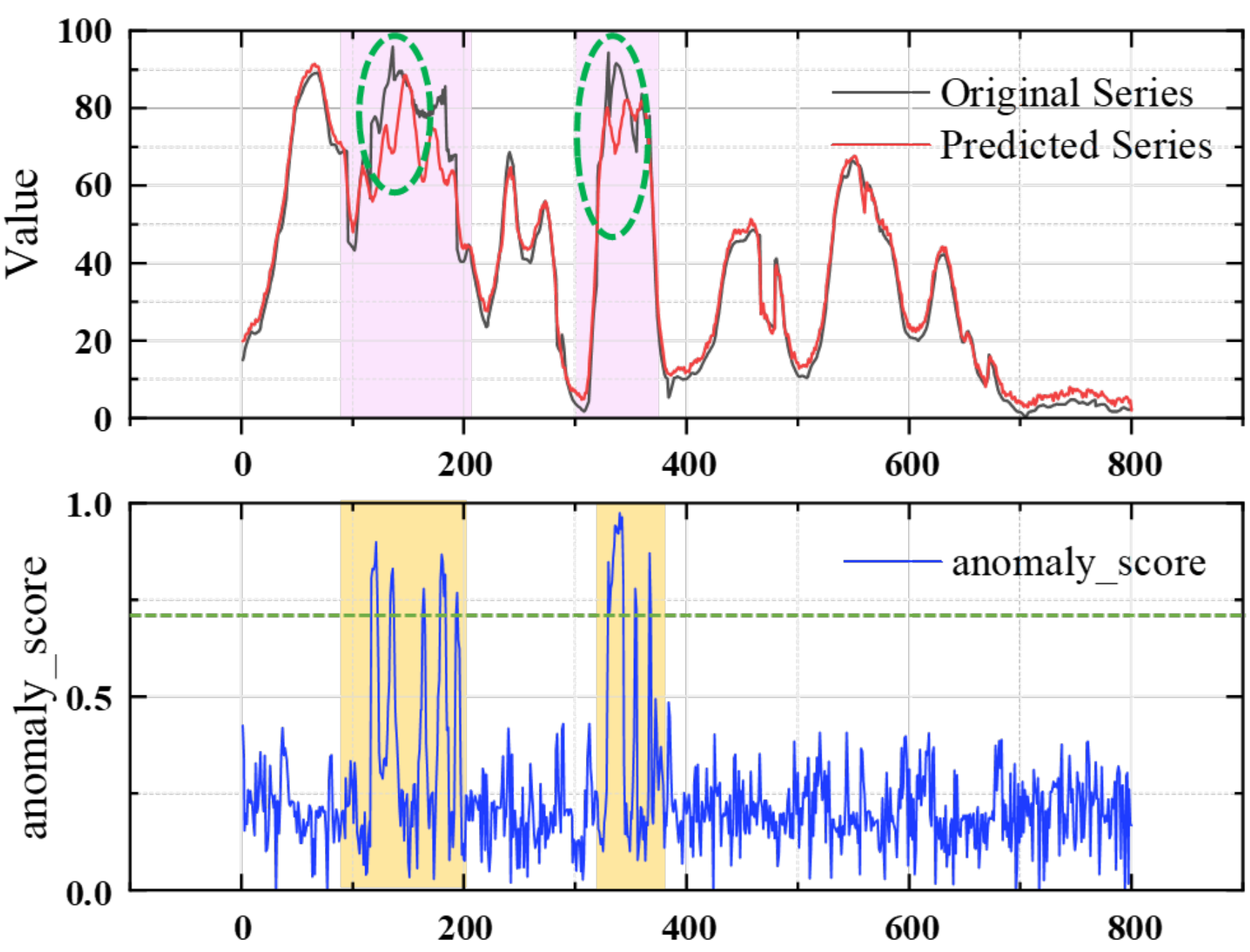

**Figure 5: Visualization of Anomaly Detection Based on Wind Power SCADA**

**Data.**

Figure 5 illustrates the anomaly detection results of a segment of SCADA data from the WTD dataset, where purple and yellow colors respectively mark the actual anomaly segments and the anomaly segments predicted by the model, and green dashed circles indicate the locations with significant reconstruction differences. The numerical range of anomaly scores enables effective identification and marking of outliers. The results demonstrate that TransGAN-WT can match anomaly patterns, accurately learning and reproducing the data features and temporal patterns under normal operating conditions of wind turbines. This validates its high reliability and precision in anomaly state identification.

## 5. Conclusion

This paper proposes a novel neural network structure, Trans GAN-WT, which can effectively capture the correlation dependencies in the multivariate time series data of wind turbines and alleviate the leakage detection problem of small deviation anomaly data. First, the detection accuracy of slight deviation data anomalies is increased by amplifying the reconstruction error. At the same time, the multimodal features are extracted using autoregressive inference, which enhances the stability and generalization ability of training. Secondly, through the temporal feature extraction module, the unit features of different time scales are prompted to be able to influence each other and learn synergistically, which effectively reduces the temporal redundancy existing in the temporal data. Finally, repeated, multi-group anomaly detection experiments were conducted on the actual wind turbine dataset, and we comprehensively compared Trans GAN-WT with other models. The experimental results show that Trans GAN-WT performs well in the task of wind turbine anomaly data detection compared to other baseline models, which provides some insights and revelations to be able to warn of potential safety accidents in the future effectively and ensure the reliability of the wind turbine to a certain extent, avoiding the losses caused by unplanned downtime.

The objective of our proposed TransGAN-WT is to perform time series anomaly detection for wind turbines, achieving early fault warning by identifying abnormal patterns in operational states. In future work, we will fully leverage the model's advantages in capturing small-deviation anomalies and multimodal features to establish a scientific anomaly severity classification system. We plan to integrate the current anomaly detection framework with degradation modeling techniques by introducing physical degradation models or deep learning-based remaining functional life prediction modules, enabling the model to provide wind turbine remaining proper life estimation functionality based on anomaly detection, thereby constructing a more comprehensive predictive maintenance solution. Additionally, we plan to introduce physics-guided regularization terms into the model's loss function to make the learning process more consistent with the actual operational mechanisms of wind turbines, further enhancing the model's interpretability and generalization capability.

**Declaration of conflicting interest:** The authors declare no conflicts of interest.
**Funding**: The authors did not receive support from any organization for the submitted work.
**Ethical approval and informed consent statements:** Not applicable.
**Data availability statements:** The datasets used and/or analysed during the current study are available from the corresponding author on reasonable request.

## REFERENCES

[1] Zhu Q, Chen J, Li F. A comprehensive analysis of China's regional energy and environment efficiency from supply chain perspective[J]. International Journal of Logistics Research and Applications, 2022, 25(4-5): 709-724.
[2] Ajanaku B A, Strager M P, Collins A R. GIS-based multi-criteria decision analysis of utility-scale wind farm site suitability in West Virginia[J]. GeoJournal, 2022, 87(5): 3735-3757.
[3] Nurseda Y. Yürüşen, Paul N. Rowley, Simon J. Watson, Julio J. Melero. Automated wind turbine maintenance scheduling[J]. Reliability Engineering and System Safety, 2020, 200:106965
[4] Wang Y, Zou R, Liu F, et al. A review of wind speed and wind power forecasting with deep neural networks. Applied Energy 2021; 304:117766.
[5] Zhang C, Hu D, Yang T. Anomaly detection and diagnosis for wind turbines using long short-term memory-based stacked denoising autoencoders and XGBoost. Reliab Eng Syst Saf 2022; 222:108445.
[6] Heydari A, Garcia DA, Fekih A, et al. A hybrid intelligent model for the condition monitoring and diagnostics of wind turbines gearbox. IEEE Access 2021;9: 89878–90.
[7] MA Kangyuan, LI Zhonghu, WANG Jinming, et al. Wind turbine fault detection based on XGBoost model[J]. Journal of Inner Mongolia University of Science and Technology, 2023, 42 (04): 349-353. DOI: 10.16559/j.cnki. 2095-2295.2023.04.009
[8] Audibert J, Michiardi P, Guyard F, et al. Usad: Unsupervised anomaly detection on multivariate time series. In: Proceedings of the 26th ACM SIGKDD international conference on knowledge discovery & data mining; 2020. p. 3395–404.
[9] ZHENG L, HU W, MIN Y. Raw wind data preprocessing: a data-mining approach[J]. IEEE Transactions on Sustainable Energy, 2015, 6(1): 11-19.
[10] ZHANG Fan, LIU Deshun, DAI Juchuan, et al. A method for recognizing the operating state of wind turbines based on the relationship of SCADA parameters[J]. Journal of Mechanical Engineering, 2019, 55(04):1-9.
[11] WANG Xin, WANG Zhengxia. Wind turbine wind speed-power data cleaning based on improved bin algorithm[J]. Chinese Journal of Intelligent Science and Technology, 2020, 2(01):62-71.
[12] A P M, A B B, B B G, et al. Real-time accurate detection of wind turbine downtime - An Irish perspective[J]. Renewable Energy, 2021,2-20.
[13] JIA Ziwen, GU Yujiong. Analysis of wind turbine gearbox operation state

based on data mining[J]. China Mechanical Engineering, 2018, 29(06):650-658.
[14] CHEN Junsheng, LI Jian, CHEN Weigen, et al. Self-coding method of wind turbine state anomaly detection using sliding window and multiple additive noise ratio stack noise reduction[J]. Transactions of China Electrotechnical Society, 2020, 35(02):345-358. DOI:10.19595/j.cnki.1000-6753.tces.181577.
[15] YU Peng, CHENG Yan, XING Jiawei, et al. Research on data-driven energy efficiency state detection and early warning of wind turbines[J]. Shandong Electric Power,2024,51(10):18-30.
[16] WANG Bote, WANG Qing, LIU Qiang, et al. A self-supervised anomaly identification method for wind turbine blades based on multi-channel vibration principal element features[J]. Journal of Zhejiang University(Engineering Science),2025,59(03):653-660.
[17] Chen Xiukang Intelligent determination method for abnormal state of SCADA data of wind turbine based on time-varying hidden Markov model [J]. Computer Measurement & Control,1-9[2025-06-04]. DOI:10.16526/j.cnki.11-4762/tp.2025.07.030.
[18] Li T, Wang Rongxi, Gao Jianmin. Method of abnormal data identification for wind turbine data acquisition and monitoring system []]. Journal of Xi'an Jiaotong University,2024,58(3):106-116.
[19] Kini, K.R.; Harrou, F.; Madakyaru, M.; Sun, Y. Enhancing Wind Turbine Performance: Statistical Detection of Sensor Faults Based on Improved Dynamic Independent Component Analysis. Energies 2023, 16, 5793.
[20] F. Harrou, R. K. Kini, M. Madakyaru, and Y. Sun, "Uncovering sensor faults in wind turbines: An improved multivariate statistical approach for condition monitoring using SCADA data," Engineering Applications of Artificial Intelligence, vol. 123, p. 106225, 2023.
[21] Lu S, Xu Q, Jiang C, et al. Probabilistic load forecasting with a non-crossing sparse-group Lasso-quantile regression deep neural network[J]. Energy, 2022, 242: 122955.
[22] Xu Q, Lu S, Zhai Z, et al. Adaptive fault detection in wind turbine via RF and CUSUM[J]. IET Renewable Power Generation, 2020, 14(10): 1789-1796.
[23] Guo B, Song L, Zheng T, et al. A Comparative Evaluation of SOM-based Anomaly Detection Methods for Multivariate Data[C]// 2019 Prognostics and System Health Management Conference (PHM-Qingdao). IEEE, 2019:25-27
[24] H. Wang, H. Xie, S. Liu, S. Song and W. Han, "A Multi-View Spatio-Temporal Feature Fusion Approach for Wind Turbine Condition Monitoring Based on SCADA Data," in IEEE Access, vol. 12, pp. 43948-43957, 2024.
[25] A. Zhu, Q. Zhao, T. Yang, L. Zhou, and B. Zeng, "Condition monitoring of wind turbine based on deep learning networks and kernel principal component analysis," Comput. Electr. Eng., vol. 105, Jan. 2023, Art. no. 108538.
[26] LIU Qingxiu, MA Hongzhan, CHU Xuening, et al. Wind turbine performance evaluation and anomaly detection based on long and short-term memory-self-coding neural network[J]. Computer Integrated Manufacturing Systems, 2019, 25(12):3209-3219. DOI:10.13196/j.cims.2019.12.022.
[27] Y. Jiang, S. Chang, and Z. Wang, "TransGAN: Two Pure Transformers Can

Make One Strong GAN, and That Can Scale Up," in Advances in Neural Information Processing Systems (NeurIPS), 2021,162:23-35.

[28] Q. Wen, T. Zhou, C. Zhang, W. Chen, Z. Ma, J. Yan, and L. Sun, "Transformers in time series: a survey," in Proceedings of the Thirty-Second International Joint Conference on Artificial Intelligence (IJCAI-23), 2023, pp. 6778-6786.

[29] Xiangjun Z, Lingqin X, Chen F, et al. An improved conditional generative adversarial method for wind turbine gearbox fault diagnosis with imbalance data[J]. IEEE Sensors Journal, 2024, 24(23): 38727-38739.

[30] HUANG Y, FU Z. Enhanced time series predictability with well-defined structures[J]. Theoretical and Applied Climatology, 2019, 138: 373-385.

[31] NIU Z, ZHONG G, YU H. A review on the attention mechanism of deep learning[J]. Neurocomputing, 2021, 452: 48-62.

[32] GOODFELLOW I, POUGET-ABADIE J, MIRZA M, et al. Generative adversarial nets[J]. Advances in neural information processing systems, 2014, 10, 27-35.

[33] Alex J. Smola†,Bernhard Scho¨ lkopf‡.A tutorial on support vector regression[J].Statistics and Computing, 2004, 14(3):199-222.

[34] HUNDMAN K, CONSTANTINOU V, LAPORTE C, et al. Detecting spacecraft anomalies using lstms and nonparametric dynamic thresholding[C]//Proceedings of the 24th ACM SIGKDD international conference on knowledge discovery & data mining. New York, NY, United States: Association for Computing Machinery, 2018: 387-395.

[35] LI D, CHEN D, JIN B, et al. MAD-GAN: Multivariate anomaly detection for time series data with generative adversarial networks[C]//International conference on artificial neural networks. Cham: Springer International Publishing, 2019: 703-716.

[36] ZHAO H, WANG Y, DUAN J, et al. Multivariate time-series anomaly detection via graph attention network[C]//2020 IEEE International Conference on Data Mining (ICDM). Sorrento, Italy: IEEE, 2020: 841-850.

[37] DENG A, HOOI B. Graph neural network-based anomaly detection in multivariate time series[C]//Proceedings of the AAAI conference on artificial intelligence. Vancouver, British Columbia, Canada: AAAI, 2021, 35(5): 4027-4035.

[38] Mucchielli, P., Bhowmik, B., Ghosh, B., & Pakrashi, V. (2021). Real-time accurate detection of wind turbine downtime-an Irish perspective. Renewable Energy, 179, 1969-1989.

[39] TULI S, CASALE G, JENNINGS N R. TranAD: Deep Transformer Networks for Anomaly Detection in Multivariate Time Series Data[J]. Proceedings of the VLDB Endowment, 2022, 15(6): 1201-1214.